\documentclass{iaus}
\usepackage{graphicx}

\title[The Galactic Bulge: A Review] %% give here short title %%
{ The Galactic Bulge: A Review }

\author[Dante Minniti \& Manuela Zoccali]   %% give here short author list %%
{Dante Minniti
  \and Manuela Zoccali$^1$
  \thanks{FONDAP Center for Astrophysics 15010003.}}

\affiliation{$^1$Dept. of Astronomy \& Astrophysics, 
P. Universidad Catolica, Casilla 306, Santiago 22, Chile 
\break email: dante, mzoccali@astro.puc.cl}

\pubyear{2007}
\volume{245}  %% insert here IAU Symposium No.
\pagerange{119--129}
\date{Aug 31 and in revised form Aug 31}
\jname{Proceedings Galactic Bulges}
\editors{M. Bureau, L. Athanassoula, \& B. Barbuy, eds.}
\begin{document}

\maketitle

\begin{abstract}
The Milky Way is the only galaxy for which we can resolve individual stars at all 
evolutionary phases, from the Galactic center to the outskirt. The last decade, 
thanks to the advent of near IR detectors and 8 meter class telescopes, has seen
a great progress in the understanding of the Milky Way central region: the bulge.
Here we review the most recent results regarding the bulge structure, age, kinematics
and chemical composition.
These results have profound implications for the formation and evolution of the
Milky Way and of galaxies in general.
This paper provides a summary on our current understanding of the Milky Way bulge, 
intended mainly for workers on other fields. 
%Due to page constraints, some of the previous efforts that opened the path to 
%the point where we stand now are often neglected.

\end{abstract}

\firstsection % if your document starts with a section,
\section{Introduction: The Questions}

\begin{figure}
\includegraphics[height=5.5in,width=5.3in,angle=00]{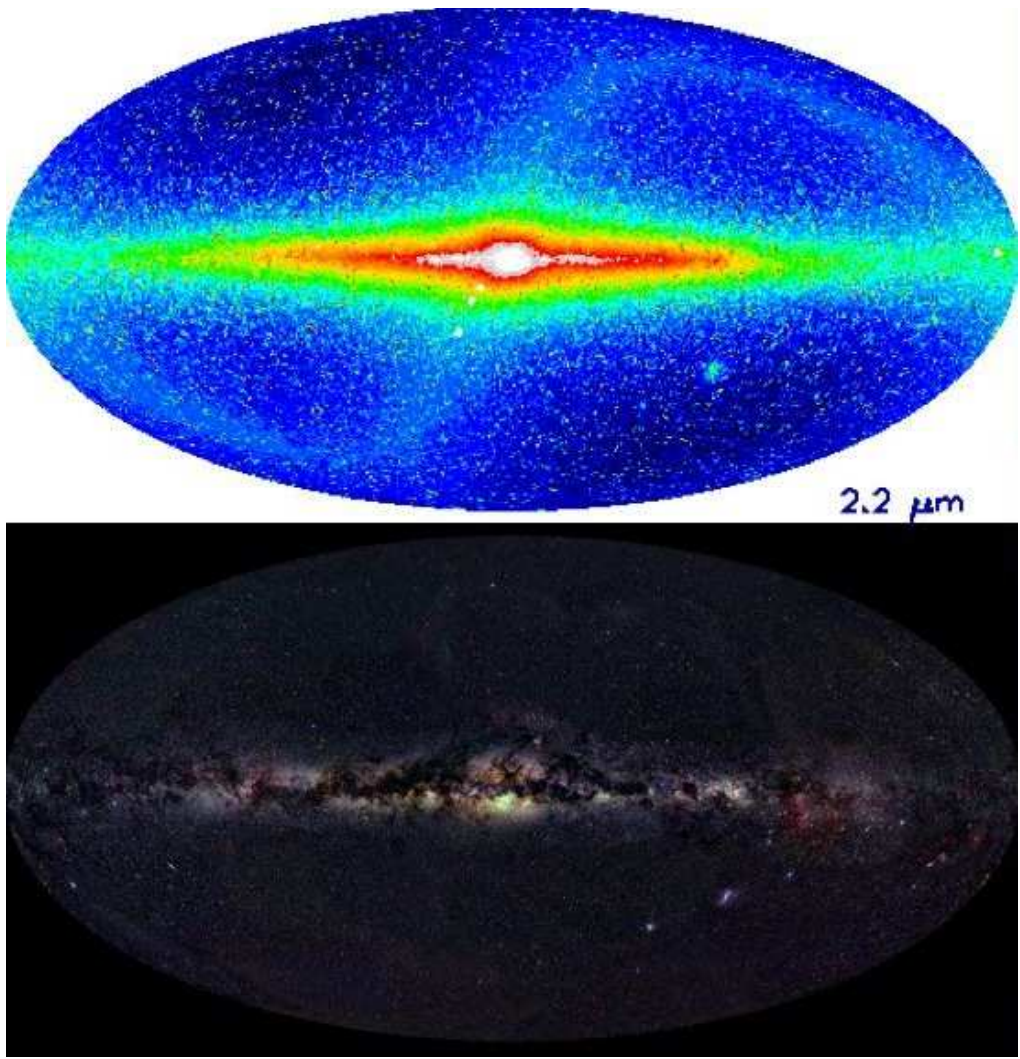}
\caption{
Top: Near-IR COBE-DIRBE map of the whole sky (Dwek et al. 1995).
Bottom: Optical map for comparison (Copyright Axel Mallenhoff 2001). 
These sky maps illustrate
why astronomers did not realize before the importance of the bulge.}
\label{maps}
\end{figure}

How did the Milky Way form?
Decades ago, the Galactic formation scenarios focused on the disk and the 
halo populations, because these were the components that astronomers 
knew something about. For example, in the classic works of Eggen et al. (1962) or 
Searle \& Zinn (1978) the bulge was not mentioned. All-sky optical maps (Fig.~\ref{maps})
did not show a clear, separate component in the inner Milky Way.

Yet upon looking at the new DIRBE or 2MASS near-infrared maps of the whole sky, 
it is evident that the Milky Way is a spiral galaxy with a peanut-shape bulge. 
%and that we live almost at the edge of the plane. 
The simultaneous observation
that bulge stars were mainly old made it clear that to answer this question the
attention had to shift towards what seems to be the first massive component to
be formed in the Milky Way.
This is a Copernican-like revolution on Galactic scales, and it is happening now! 
We see our Galaxy as if it were an external galaxy for the first time, and are 
fortunate to have such new perspective, and also to be able to provide specific 
answers to the many questions regarding its formation.

%By the way, this serves to recognize the 2MASS experiment
%as an important legacy for galactic astronomy. The 2MASS point source 
%catalogue is public, and gives JHK magnitudes and positions for point sources 
%across the Galaxy, allowing (revolutionizing) studies of stellar populations, 
%reddening, galactic structure, etc. (Cutri et al. 2003). 

Note that we have only one object to study: our Galaxy. This is a fundamental 
limitation because we have to get the answers right. 
%In this respect this field is similar to Cosmology, with only one Universe 
%to study, as elegantly put by Disney (2000). 
The advantage in the case of our Galaxy, of course, is that we 
can study the subject in detail: in no other galaxies the fundamental problems 
of Astronomy can be surpassed. Basically, these fundamental problems are: 
(1) The distance problem: we can resolve the components of the Milky 
Way into stars, and study them in situ, obtaining 3-D positions and motions,
and detailed chemical compositions of individual stars; and 
(2) The timescale problem: we see an instant snapshot, 
and must infer histories from that.
In our galaxy we can separate and date the components,
using the main-sequence turn-offs 
of different stellar populations and clusters. 
%In distant galaxies we have the limitation that "No evolutionary population 
%synthesis code is perfect" (Renzini 2007).

In spite of a small community working on the Galactic bulge, 
%one should not underscore the importance of this field.  The answers that this community 
%gives must be taken into account by the wider astronomical community. And it so happens that
a revolution has occurred in this field in the last decade,
with great progress in comparison with other fields of Astronomy.
For the first time, we have the answers to important questions such as:

{\bf 1.} Is the bulge a different Galactic component? 

{\bf 2.} When did the Galactic bulge form? 

{\bf 3.} How did it form?

{\bf 4.} Is there a radial gradient in the bulge? 

{\bf 5.} Are there globular clusters associated with the bulge? 

{\bf 6.} Are there planets in the bulge?

The 8m class telescopes were built last decade, aiming 
to answer these questions. The proposed options were endless (bulge
formation from secular evolution of the disk, from the halo, in a single
burst, in several episodes, by slow accretion of smaller sub-units, etc).
The evidence and the answers described below serve  as a  basis for
understanding   more    distant galaxies    which  cannot  be  studied
(dissected) in similar detail.

\section{Structure of the Galactic Bulge}

The bulge is elongated, its  barred structure was
clearly shown by  the DIRBE IR  maps (Dwek  et  al. 1995),  and by the
clump  giants (Stanek et  al. 1994).   This  structure converges to  a
barred bulge of axes 1:0.35:0.26, inclined  by 25 degrees with respect
to the  line  of sight,  with  the  nearest side  located at  positive
longitudes (e.g. Rattenbury et al. 2007b).
%  {\bf MZ: do we know the scale lenght? This depends on Ro, no concensus}

Much of the  progress on   the bulge  structure has   come  out as   a
byproduct  of  the   microlensing experiments:  from  the microlensing
events,  from the  color-magnitude diagrams   (CMDs), and from   the
variable stars. These databases  are treasures for the whole community
to exploit.

While stellar populations  trace light  in galaxies, the  microlensing
optical  depth  is directly  sensitive to   the mass. The microlensing
experiments have concentrated on bulge fields, discovering hundreds of
events.  The  MACHO clump giants give  a  bulge optical  depth $\tau =
2.17\times 10^{-6}$ at $l, b = 1.5, -2.7$  (Popowski et al. 2005). The
EROS  and OGLE bulge results are  in agreement (Hamadache et al. 2006,
Sumi et  al. 2006). The modeling of the inner barred mass distribution
appears now in agreement (Gerhard 2005), with a bulge total mass of
$1.6\times10^{10} M_\odot$. 

\begin{figure}
\includegraphics[height=2.0in,width=3in,angle=00]{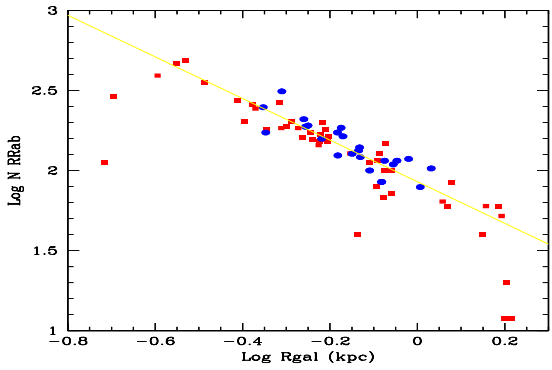}
\caption{
Number density of RR Lyrae type ab in the inner bulge $vs$ Galactocentric
distance from Alcock et al. (1998, circles) and Collinge et al. (2006,
squares).  The line is a spherical distribution with $\rho \propto r^{-2.3}$.  
The  numbers per sqdeg have been normalized to Baade's window. Even though the 
innermost fields are still likely to be incomplete, the known RR Lyrae
clearly show a very concentrated  distribution.
This figure  shows that old and  metal-poor populations are also
present in  the  inner bulge  regions.
%  which have   a clear composite
%population   dominated   by    the  metal-rich   stellar    component.
}
\label{rrlyr}
\end{figure}

However, there are still puzzles.
Thousands of RR Lyrae have been found in the bulge. As classical tracers of
old metal-poor populations, they may  belong to the inner halo
rather than to the bulge. At Baade's window the RR Lyrae have mean
$[Fe/H]=-1.0$ (Walker \& Terndrup 1991). In fact, their distribution is
not so clearly  elongated as that  of  bulge clump giants,  although a
small barred structure  may be  present (Alcock et al.  1998, Collinge 
et al. 2006). Nevertheless, the strong central concentration of the RR
Lyrae distribution (Fig.~\ref{rrlyr}) illustrate the
composite nature of the inner stellar population.
An interesting possibility is that the metal-poor stars and globular 
clusters located in the inner bulge
might be the oldest populations in the Galaxy (e.g. Mackey \& van den Bergh 2005).

There also  appears to  be a small  bar  (scale $\sim 600$ pc)  in the
central region (Alard 2001, 
Niyishama  et al.  2006), and  a  larger bar structure
confined to the plane (e.g. Cole \& Weinberg   2002, Lopez-Corredoira 
et al. 2007).  Indeed, much remains  to  be explored regarding the
structure  of the inner bulge.   And  more considering that the galaxy
morphology  might be  a transient  phenomenon (Steinmetz 2007),
and that bars might appear and disappear (Combes 2007).

\section{Kinematics of the Galactic Bulge}

\begin{figure}
\includegraphics[height=2.0in,width=5in,angle=00]{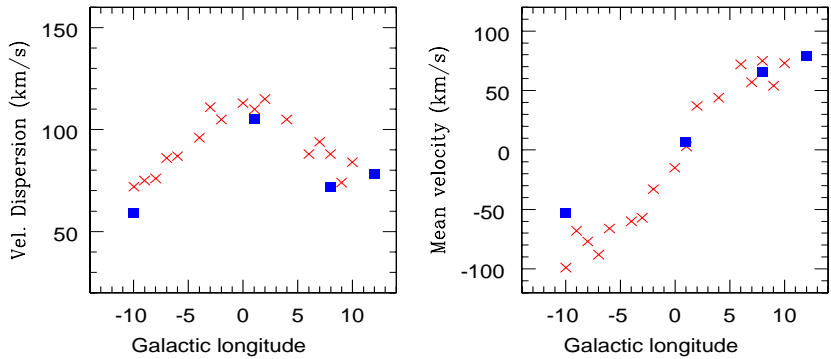}
\caption{ Bulge  rotation from radial    velocities of K-giants   as
measured by Minniti (1996), compared with the measurements of M-giants
by Rich et al. (2007) (crosses).  The latter have been corrected
for  the  Solar motion around the   Galaxy,  which was  not taken into
account in the original paper.}
\label{rotation}
\end{figure}

A great effort  in the  90's was  devoted to  the  measurement of  the
kinematics of  bulge  K-giants , representative of   all  bulge
populations    (as  all giants go  through   this  stage regardless of
metallicity),     using      radial   velocities.     As     shown  in
Fig.~\ref{rotation},  the bulge is  rotating, with  a peak rotation of
about 75 km/s (Minniti et al.  1992, Harding \& Morrison 1993, Minniti
1996, Ibata et al. 1995, Beaulieu et al. 2006, Rich et  al. 2007).  It
also has a large velocity dispersion (Terndrup et al. 1995, Minniti et
al.  1996, Ibata  et  al.  1995),  that decreases with  Galactocentric
distance.

\begin{figure}
\includegraphics[height=2.5in,width=3in,angle=00]{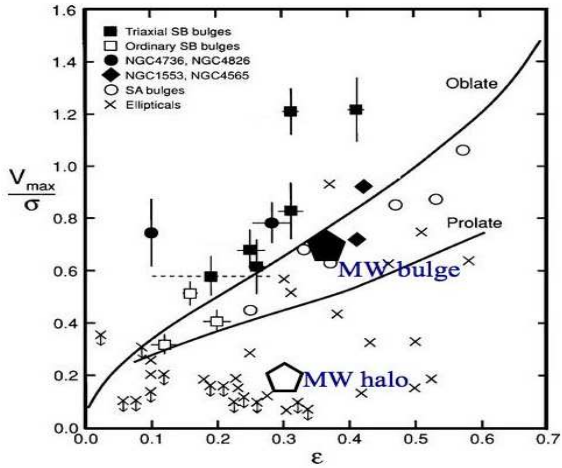}
  \caption{
Milky Way bulge and halo position in the Binney (1978) diagram adapted 
from Minniti (1996) and Kormendy \&  Kennicutt (2004).
}\label{Vsigma}
\end{figure}

The bulge kinematics are intermediate between a purely rotating system
such as  the Milky Way  disk and a  hot, non-rotating  system like the
Milky  Way halo, that is  supported  by velocity  dispersion.  The  so
called  $V_{max}/\sigma$   diagram  (Binney  1978)   is a  measure   of  how
kinematically hot is  a  particular stellar system.  This  diagram has
recently  been  adapted by  Kormendy \&  Kennicutt (2004) for external
bulges.  Fig.~\ref{Vsigma} shows  an update of  the bulge position  in
this diagram, with  $(V_{max}/\sigma)_{MWB}=0.67$ for the metal-rich bulge component,
indicating that this system is kinematically  hotter than the Milky Way
disk  (with $(V_{max}/\sigma)_{MWD} \sim 2.2$ using the $\sigma_{MWD}$ measured by 
Morrison et al. 1990), but colder than the halo with $(V_{max}/\sigma)_{MWH}\sim 0.15$
(Minniti 1996, Battaglia et al. 2005).

The  proper motions were measured in   Baade's window by Spaenhauer et
al. (1992), confirming the high  velocity dispersion of M-giants. More
recent    proper-motion measurements from    the  ground and space are
following their steps (Kuijken \& Rich 2002, Soto  et al. 2007, Vieira
et al. 2007, Rattenbury et al. 2007a), and this  is a field where a lot
of progress is expected: we need orbits of bulge stars, to be able to
distinguish between a possible inner disk/halo component and the true
bulge.

\section{The Bulge Globular Cluster System}

The metal-rich globular clusters in  the central regions of the  Milky
Way share the kinematics, spatial distribution, and composition of the
bulge   field stars (Minniti  1995,  Barbuy  et  al. 1998, C\^ot\'e
1999,  Bica  et al. 2006).   This  realization was very  important
because it allowed  us to trace the bulge  stellar population by means
of its globular    clusters.   Typical bulge  globular clusters   like
NGC6553,  NGC6528, NGC6441,  have  been  well  studied,  with accurate
distances, reddenings,  ages, individual star memberships and detailed
chemical  compositions (e.g. Barbuy et  al.  2005, 
Zoccali et al. 2004, Alves-Brito et al. 2006), and serve now as 
standards for bulge studies. The detailed chemical compositions of the
old metal-poor globular clusters present in the bulge are being
measured as well (Barbuy et al. 2006, 2007).

%\section{Stellar Populations of the Galactic Bulge}

\section{The Age of the Galactic Bulge}

The bulk of the metal-rich stellar population of the Galactic bulge is old,
with an age  of $t=10\pm 2.5$ Gyr   (Ortolani et al. 1995,  Zoccali et
al. 2003).   This  result has been   obtained by direct  comparison of
bulge field stars with metal-rich globular clusters.

\begin{figure}
\includegraphics[height=5in,width=5in,angle=00]{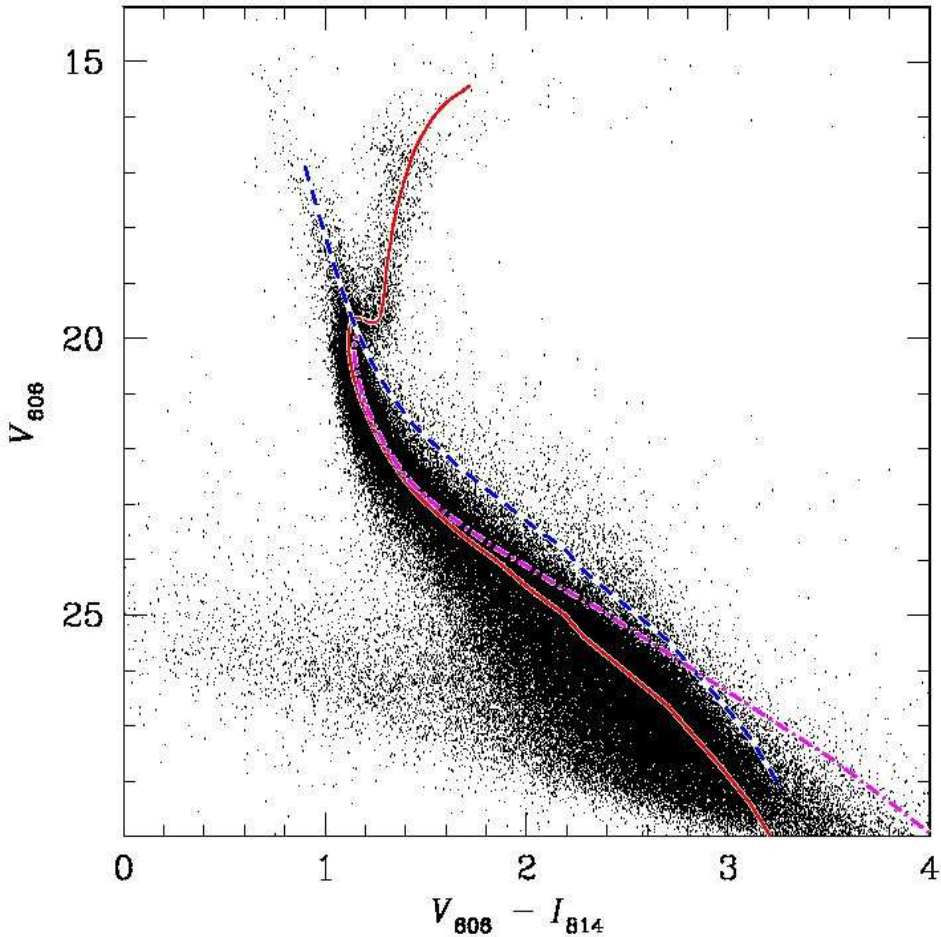}
\caption{
The deepest bulge CMD from $86,106$ and  $89,835$ sec in the F606W (V)
and  F814W (I)   bands,  obtained  with the  ACS    at the HST on    a
$202""\times    202"$  field   at   $l, b  =  1.25,  -2.65$   (Sahu et
al. 2006).  About 245,000  stars are plotted  down  to $V=29$. A Solar
metallicity isochrone of 10 Gyr is plotted with  the red line, fitting
the bulk of the  bulge population. The foreground  disk is fit with an
unevolved main  sequence  shown by  the   blue dashed  line.  A  super
metal-rich isochrone with $[Fe/H]=+0.5$ is also shown with the magenta
dot-dashed line.}
\label{cmd}
\end{figure}

The deepest CMD of the bulge to date was  obtained with the ACS at HST
by  Sahu  et al.  (2006).  This diagram,  shown in  Fig.~\ref{cmd}, is
consistent with such an  old age. These data  were acquired during  an
HST  large  programme to find  planetary transits   in  the bulge: the
SWEEPS project  (Sagittarius   Window  Eclipsing  Extrasolar   Planets
Search).   Sahu  et  al. (2006) found  16   transiting  planetary size
objects, two of which  were confirmed to  be real planets using radial
velocities.

In  spite of that, there  are tracers of  a younger population such as
OH/IR   stars, bright AGB  variables, etc.,  that  appear to be mostly
confined to the  Galactic plane (e.g. van Loon  et al. 2003). The bulk
of the bulge, however, is old. 

\section{The Chemical Composition of the Galactic Bulge}

In principle, the measurement  of the age  distribution of bulge stars
should allow us to derive its star formation  rate, and thus to answer
at least the question on   when (if not  how) did  the bulge form.  In
practice, the  disk contamination right  on top of  the bulge
turnoff, coupled with     the  metallicity, distance  and    reddening
dispersions did not permit the dating of the bulge to better than $\pm
2.5$ Gyr.  Complementary observation of the  metallicity distribution
and  the detailed element ratios   allowed to constrain the  formation
timescale, the star formation  rate, initial mass function, the infall
%(and its timescale) 
of primordial material, and the possible occurrence
of accretion of smaller sub-units.

\subsection{The Metallicity of the Galactic Bulge}

The bulge metallicity is far from that  of a simple stellar population
like a globular cluster, and therefore a large number of stars need to
be measured in order to sample the whole metallicity range. Being high
resolution spectroscopy of a large number of stars prohibitive, in the
past people
would  determine the metallicity  distribution via low resolution
spectra for a few hundred stars  (e.g. Sadler et  al. 1996, Ramirez et
al. 2000) and  sometimes obtaining high resolution  spectra only for a
few    dozen  stars, used  as  calibrators  (McWilliam  \&  Rich 1994,
Fulbright, McWilliam \&   Rich 2006). They would  all   agree that the
Galactic bulge  was metal rich, but   there was contradictory evidence
regarding the mean, the spread and the shape of the distribution.

\begin{figure}
\includegraphics[height=2.0in,width=3in,angle=00]{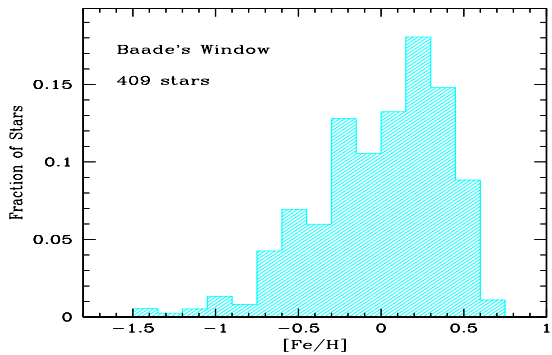}
\caption{
Metallicity distribution of 
%204 K-giants in the  bulge from Zoccali et al. (2007)
409 RGB and clump K-giants in the  bulge from Zoccali et al. (2007) and Lecureur et al. (2007b),
all measured for the first time with high-resolution spectra. 
This is in a  firm  metallicity scale  as they observed simultaneously 
a dozen stars of known globular clusters located in the bulge.}
\label{MDF}
\end{figure}

Zoccali et al.  (2007) and Lecureur et al. (2007b) 
have now obtained  for the first time the bulge
metallicity distribution from a  sample of $\sim$  1000 K-giants  in 4
bulge  windows,  {\it all}   observed  with  high dispersion   spectra
($R>20000$), with FLAMES at VLT. The resulting metallicity distribution
for Baade's Window is shown in Fig.~\ref{MDF}.  Its shape is well
reproduced  by the most  recent  chemical evolution models (Ballero et
al. 2006) assuming a very high star formation efficiency, short infall
timescale, and a rather top-heavy  initial mass function (as  measured
in the bulge by Zoccali et al. 2000, for $1.0<M/M_\odot <0.15$).

\subsection{The Detailed Element Abundances of the Galactic Bulge}

\begin{figure}
\includegraphics[height=5.5in,width=2.0in,angle=270]{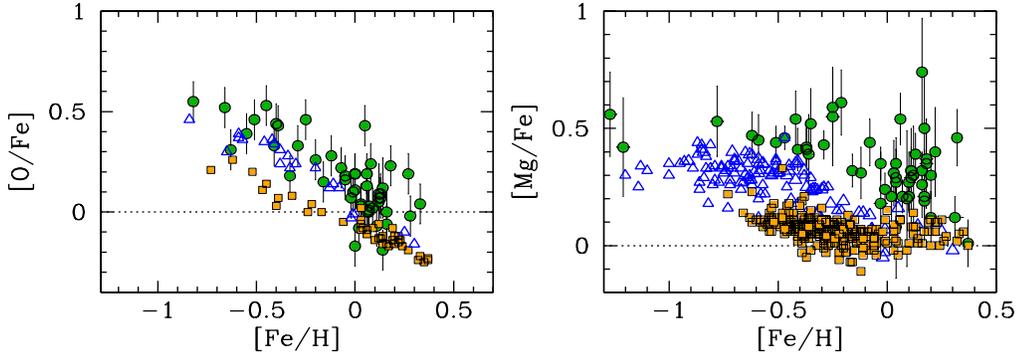}
\caption{  Oxygen and  magnesium over iron  ratios  as measured from
high dispersion spectra  of   bulge K  giants (Zoccali  et   al. 2007,
Lecureur et al. 2007a).  Green circles  with error bars are bulge stars,
compared  with  thick (blue triangles)  and  thin (red squares) disk
stars.  These results  clearly indicate that   the bulge  formed  as a
separate component by a rapid chemical enrichment.  }\label{alphas}
\end{figure}

Different teams were pursuing the detailed element abundances of bulge
giants using high dispersion  spectrographs, following the pioneering
effort by McWilliam \& Rich (1994).  Recently,  Rich \& Origlia (2005)
found  that $\alpha$-elements are enhanced  to  [O/Fe]=+0.4 for 13 giants
within a  narrow   metallicity range  around   [Fe/H]=-0.2. Later  on,
Zoccali et al. (2006) and  Lecureur et al. (2007a) measured oxygen,
magnesium, sodium and aluminum in a sample  of 50 K-giants with [Fe/H]
covering a wide metallicity range,
from $-0.8$   to $+0.3$ using the  UVES  spectrograph at  the VLT. The
result is shown in Fig.~\ref{alphas},  where it is evident that  bulge
stars have larger  [O/Fe] and  [Mg/Fe]  than both thin  and thick disk
stars.   This is the  signature of  a chemical  enrichment  by massive
stars, progenitors of SNII, with  little or no contribution from 
SNIa, and thus of  a shorter  formation  timescale of the  bulge
with respect  of  both disk  components.   In  this  sense, the  bulge
(including   its  globular clusters)   is  the  most  extreme Galactic
population.

These  results  agree with the predictions   of recent bulge formation
models  all assuming  a fast  formation  at early   epochs (Ballero et
al. 2006, Immeli et al.  2004).  Most importantly, they were confirmed
by independent teams: Fulbright  et al. (2007)  analyzed 27 K  giants,
and Cunha \& Smith (2007) observed 7  giants, all with enhanced Oxygen
with respect to the other Galactic components, at all metallicities.

\section{The Stellar Population Gradient of the Galactic Bulge}

Is there  a   stellar   population   gradient  in   the    bulge?  The
CMDs  of the microlensing  surveys  that mapped a
large   fraction of the   bulge   showed significant  differences with
increasing  distance from  the  center and  from the  plane.  The most
conspicuous effect is  that  the  red  giant  branch gets   bluer  and
narrower  moving  away from the  Galactic center  (Fig. 7).  This
observed gradient  can be due to:  the different  contributions of the
bulge   $vs$ disk   $vs$ halo, reddening/extinction   inhomogeneities,
crowding, a variation in  chemical composition, age differences, or  a
combination of any of the above.

Several  pieces of evidence  pointed  towards  a metallicity  gradient
(Minniti   et al. 1995), however  a  very large spread existed between
different authors' measurements   on  the same  fields.   Furthermore,
recent analysis (Ramirez et  al. 2000; Rich,  Origlia \& Valenti 2007)
suggested  the absence of  a  gradient  in  the inner  region ($b=1-4$
degrees, i.e., inside Baade's Window). The final  answer comes from 
the homogeneous iron abundances from high dispersion spectroscopy  of 
nearly 1000 K-giants (Zoccali et al. 2007 and Lecureur et al. 2007b). 
The bulge  has a   metallicity gradient, with  the
metallicity  decreasing by $0.25$   dex along the Galactic minor  axis
between $b=-4$ and $b=-12$ deg.

%\begin{figure}
%\includegraphics[height=3in,width=4in,angle=00]{minniti_fig18.ps}
%\caption{
%The metallicity gradient measured along the bulge minor axis, using high 
%dispersion spectroscopy of K giants (Zoccali et al. 2007). Not only the 
%mean metallicity changes, but also the shape of the metallicity distribution. Black gaussians
%show the position (Nordstr\"om et al. 2004) and relative fraction (from the
%Besan\c on model, Robin et al. 2003) of thick and thin disk contaminants.}
%\label{gradient}
%\end{figure}

\begin{figure}
\includegraphics[height=3in,width=5.5in,angle=00]{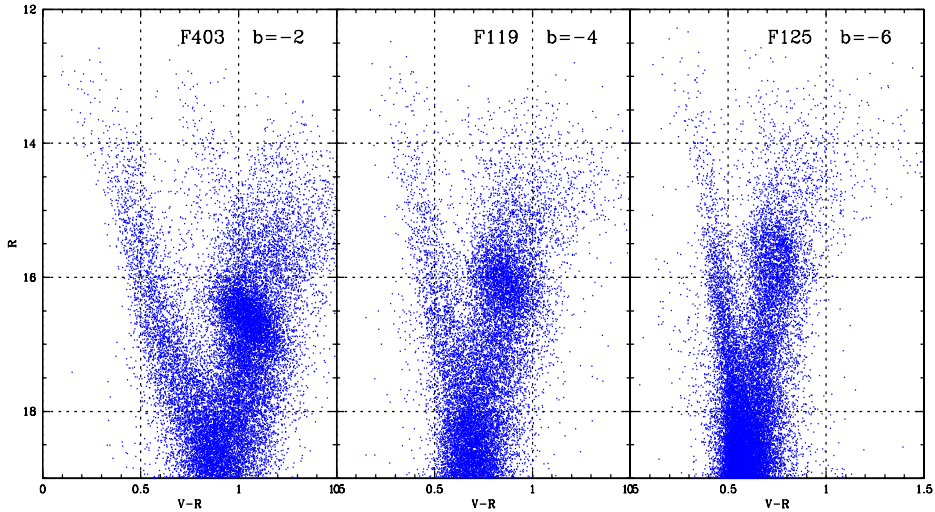}
\caption{
CMDs for three MACHO fields along the bulge minor axis.
The change in the stellar population is evident, 
with the RGB getting bluer and narrower away from the Galactic center. This
cannot only be due to reddening, and is explained by the metallicity gradient 
measured 
%along the bulge minor axis, 
using high dispersion spectroscopy of K giants (Zoccali et al. 2007,
Lecureur et al. 2007b). 
}
\label{gradient}
\end{figure}

\section{Conclusions: The Answers}

For the first time, we have the answers to the following basic questions:

{\bf 1.} Is the bulge a different component? Yes, based on all the evidence available,
the bulge is a distinct Galactic component, with different kinematics and 
compositions from the thin disk, the thick disk and the halo.

{\bf 2.} How did the Galactic bulge form? It formed on a
short timescale ($\sim 1$ Gyr), as demonstrated by the $\alpha$-element enrichment. 
Despite the presence of the bar, models like a bulge formation via secular 
evolution of the disk can be firmly excluded.
%In addition, no evidence of the presence of subgroups has been found so far.

{\bf 3.} What is the age of the bulge? 
The bulk of the stellar population is $\sim 10$ Gyr old. However,
there are traces of a small fractions of intermediate-age stars, 
and of metal-poor stars. The latter might well be the oldest population in the Galaxy. 

{\bf 4.} Is there a gradient in the bulge? Yes, there is a stellar population
gradient as shown for example by the CMDs. Now it is
found that this gradient is mostly due to metallicity, which
decreases along the Galactic minor axis. 

{\bf 5.} Are there globular clusters associated with the bulge? Yes, there is 
a population of metal-rich globular clusters in the central regions 
that share the kinematics, spatial distribution, and composition of the bulge 
field stars. 

{\bf 6.} Are there planets in the bulge? Even though this question seems to belong 
to another field, another of the recent advances was the discovery of planets
in the bulge by HST. The available data suggests that giant planets
are as numerous in the bulge as they are in the Solar neighborhood.

These revolutionary advances that impact the whole of extragalactic Astronomy cannot be 
attributed to the success of a single group, but to the combined contributions 
of different teams. Where controversy was present before, today similar answers are 
given. Progress has occurred!
%And a very small community is responsible for the enormous progress. This
%it is a low-budget community, that has been living out of small resources (e.g. no 
%mega-budget projects, no dedicated telescopes or satellites).  
%
%In spite of this progress, important steps still need to be made to understand the
%Galactic bulge. And off course, we have new important questions to attack:
%
%What are the orbits of bulge stars? Proper motions becoming available will teach us. 
%
%What are the best models? Intense activity on this front is providing 
%fits to the data.
%%, specially the detailed bulge chemical composition.
%
%What is the inner bulge structure? We must explore outside low extinction windows.
%% Most programs are necessarily concentrated to samples located in a few windows of low extinction.

\section{Future Dreams}

We are now dreaming of the bulge science that would be enabled 
with the next generations of 4-8m class survey telescopes, and of 20-40 m class ELTs.
Our dreams are that
the survey telescopes like VISTA will allow us to map the whole bulge, unveiling
its entire structure and stellar populations (the VVV survey), and
that the ELTs will allow us to measure the chemical compositions of bulge
turn-off stars as function of their ages.

%\begin{acknowledgments}
%We are supported in part by FONDAP Center for Astrophysics 15010003.
%\end{acknowledgments}

%\begin{discussion}
%
%\discuss{Massey}{wondering if you have considered the intrinsic dispersion in absolute
%}
%
%\discuss{van der Hucht}{Indeed, we will be always left with 
%}
%
%\discuss{Walborn}{I think that the scatter in 
%}
%
%\discuss{van der Hucht}{As said above, there is 
%} 
%\end{discussion}

\end{document}